\documentstyle[11pt,newpasp,twoside]{article}
\def\edcomment#1{\iffalse\marginpar{\raggedright\sl#1\/}\else\relax\fi}
\reversemarginpar

\newcommand\about     {\hbox{$\sim$}}

\newcommand\RRc       {candidate RR Lyrae stars}
   
\def\vol@title{Milky Way Surveys: The Structure and Evolution of Our Galaxy}

\begin{document}
\phantom{x}
\vskip 0.1in

\title{Halo Structure Traced by SDSS RR Lyrae}

\author{\v{Z}eljko Ivezi\'{c}$^1$, Robert Lupton$^1$, David Schlegel$^1$, 
Vernesa Smol\v{c}i\'{c}$^{1,2}$, David Johnston$^3$, Jim Gunn$^1$, 
Jill Knapp$^1$, Michael Strauss$^1$, Constance Rockosi$^4$ (SDSS Collaboration)}

\affil{$^1$ Princeton University Observatory, Princeton, NJ 08544}
\affil{$^2$ University of Zagreb, Dept. of Physics, Bijeni\v{c}ka cesta 32, 
            10000 Zagreb, Croatia}
\affil{$^3$ University of Chicago, Astronomy \& Astrophysics Center, 
            5640 S. Ellis Ave., Chicago, IL 60637}
\affil{$^4$ University of Washington, Dept. of Astronomy,
            Box 351580, Seattle, WA 98195}

\begin{abstract}
We discuss the density and radial velocity distributions of over 3000 \RRc\
selected by various methods using Sloan Digital Sky Survey data for about 
1000 deg$^2$ of sky. This is more than 20 times larger sample than 
previously reported by SDSS (Ivezi\'{c} et al. 2000), and includes \RRc\ out 
to the sample limit of 100 kpc. A cutoff in the radial distribution of halo 
RR Lyrae at $\sim$50-60 kpc that was suggested by the early SDSS data appears 
to be a statistical anomaly confined to a small region ($\sim$100 deg$^2$). 
Despite the large increase in observed area, the most prominent features 
remain to be those associated with the Sgr dwarf tidal stream. We find 
multiple number density peaks along three lines of sight in the
Sgr dwarf tidal stream plane, that may indicate several perigalactic 
passages of the Sgr dwarf galaxy.
\vskip -0.3in
\end{abstract}

\section{The Sample} 

The Sloan Digital Sky Survey (SDSS) is significantly contributing to the 
studies of Galactic structure. Extensive information about the SDSS can be 
found in Azebajian et al. (2003, and references therein). Here we discuss a sample
of 3,127 candidate RR Lyrae stars selected in $\sim$1,000 deg$^2$ of sky
using color and variability information obtained from multiple SDSS imaging 
and spectrophotometric data. Details about the selection of
candidate RR Lyrae stars using SDSS can be found in Ivezi\'{c} et al. (2000, 
hereafter I00). This new sample is over 20 times larger than that analyzed 
by I00, and offers significant new clues about the Galactic halo structure. The estimated 
sample completeness is about 35\%, and its efficiency (the fraction of true 
RR Lyrae stars) is about 90\%. Additional analysis of the same sample is presented 
in a companion paper (Ivezi\'{c} et al. 2003, hereafter I03). Here we attempt 
to answer the following questions:

\begin{itemize} 
\item
Is the cutoff in the radial distribution of halo RR Lyrae at $\sim$50-60 kpc
suggested by the early SDSS data (I00) supported by the enlarged sample
discussed here?
\item
Are there additional halo substructures comparable to those associated
with the Sgr dwarf tidal stream?
\item
Is there evidence for the multiple perigalactic passages of the Sgr dwarf
galaxy? 
\end{itemize}

\begin{figure}[t]
\plotfiddle{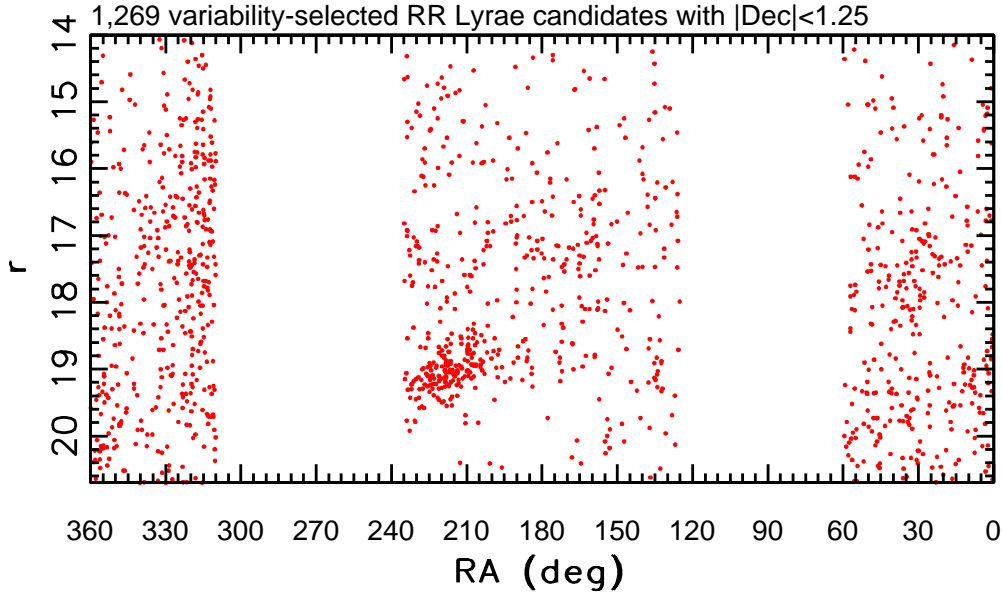}{7cm}{0}{80}{80}{-250}{-390}
\caption{The  $r$ vs. RA distribution for \about 1000 SDSS candidate RR Lyrae stars 
along the Celestial Equator ($|Dec|<1.25^\circ$).}
\end{figure}

\begin{figure}[ht]
\plotfiddle{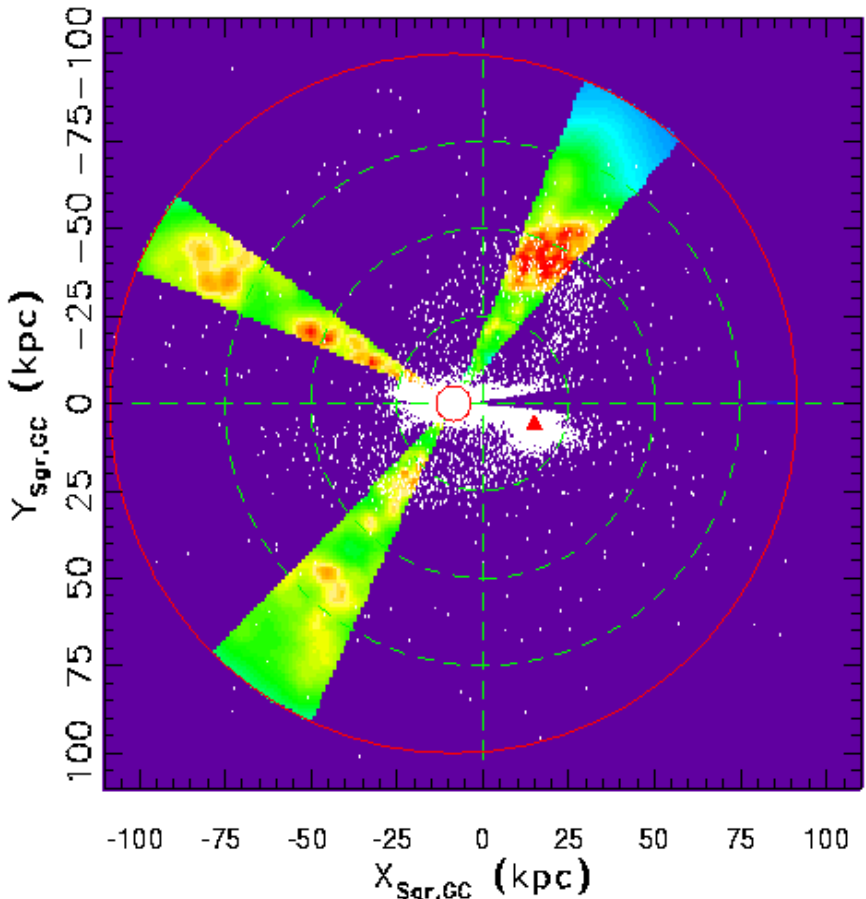}{10.8cm}{0}{62}{62}{-200}{-90}
\caption{
The number density multiplied by the cube of the galactocentric radius
(logarithmic scale with dynamic range of 1000, from light blue
to red), for 923 RR Lyrae stars within 10$^\circ$ from the Sgr 
dwarf tidal stream plane. The 2MASS M giants discussed by M03 are marked
by the white dots. The solid circles show the sample 
distance limits (5 kpc and 100 kpc), and the dashed circles, centered on
(X=0,Y=0), are added to guide the eye. The red triangle marks the Sgr dwarf core.
} 
\end{figure}

\section{   Analysis }

\subsection {The Apparent Magnitude Distribution of the Selected Candidates}

The distribution of a subsample of 1,269 \RRc\ along the Celestial Equator ($|Dec|<1.25$)
in $r$ vs. RA diagram is shown in the top panel in Figure~1. The region studied
by I00 corresponds to the range $160 < RA < 235$. As discernible from the figure,
there seems to be an abrupt decrease in the number of selected candidates with 
$r > 19.5$ in this region. However, the new data indicate that this is an
anomaly; outside that region candidates are found all the way to the sample
faint limit at $r=20.7$, including other directions on the sky not shown in 
the figure (see the companion paper, I03).

\subsection{ Halo Substructure As Traced By RR Lyrae Stars }

The distribution of selected stars shown in 
Figure~1 is very inhomogeneous. Especially prominent features are the 
clump associated with the Sgr dwarf tidal tail (leading arm) at RA \about 200-230, 
$r$ \about 19, Pal 5 globular cluster and associated tidal debris 
(RA \about 230, $r$ \about 17.4), and a clump at RA \about 180, 
$r$ \about 17. The latter is also detected by Vivas et al. (2001) 
using a complete sample of confirmed RR Lyrae stars discovered by the
QUEST survey. Another prominent feature is the so-called ``southern'' clump, 
also associated with the Sgr dwarf tidal tail (trailing arm), and discovered 
using A-colored stars by Yanny et al. (2000). We note another, previously
unreported, overdensity of fainter stars ($r$ \about 19-20, D \about 60-70 kpc) 
in approximately the same region (RA \about 30-60). The increased number of stars 
around RA \about 320 coincides with the point where the sampled part of
the Celestial Equator is closest to the Galactic center, and is an expected 
consequence of the $r^{-3}$ density distribution (Wetterer \& McGraw 1996, and 
references therein). In summary, although the sample is increased by a factor of 20,
the Sgr dwarf tidal stream remains by far the strongest detected feature.

\subsection{  The Properties of the Sgr Dwarf Tidal Stream }

About 1000 \RRc\ from the sample discussed here are found within 10$^\circ$ from 
the  Sgr dwarf tidal stream plane, as defined by Majewski et al. (2003,
hereafter M03). We use this subsample to search for evidence of multiple 
perigalactic passages, which should manifest themselves as multiple peaks in the
number density along a line of sight that is in the orbital plane.
The smoothed
distribution of these stars is shown 
in Figure~2 (the employed coordinate system is same as in Figure~11 from 
M03). 
For unsmoothed data points, and histograms of the distance and radial velocity
distributions see I03.

As evident, there are multiple peaks along each of the three lines of
sight. The strongest clump is a part of the leading arm of the Sgr dwarf tidal tail. 
The weakest features visible in this figure are still marginally statistically 
significant (2-3$\sigma$), as determined  by simulations of random samples of the 
same size. These multiple peaks suggest that 
the Sgr dwarf may have experienced multiple perigalactic passages, as predicted by 
some models (e.g. Ibata et al. 2001), but not all (Martinez-Delgado et al. 2003). 
Alternatively, comparison with an all-sky view of M giants (M03), shows by the white
dots, indicates
that perhaps some of the new clumps are part of the main trailing arm that extends
from the Sgr core through the southern Galactic hemisphere, curves back towards the 
Galactic plane and continues through the northern Galactic hemisphere (see also 
Newberg et al. 2003). The clumps at (X,Y) of (-25,25) and (-80,-40) are consistent 
with this hypothesis, but the clumps at (-45,55) and (-45,-20) remain unexplained. 
On the other hand, the structures at large radii could be fluctuations unrelated 
to the Sgr dwarf tidal stream (e.g. see Bullock et al. 2001). For example, analogous 
analysis (such as that illustrated in Fig.~2) of candidates along the Celestial Equator 
(see Fig.~1) reveals only slightly weaker overdensities that are more than 30$^\circ$ 
away from the Sgr dwarf tidal stream plane. A more detailed statistical 
analysis is required to draw definitive conclusions, and will be presented in a 
forthcoming publication.  

\vskip -0.15in
\phantom{x}

{\small 
Funding for the creation and distribution of the SDSS Archive has been provided by 
the Alfred P. Sloan Foundation, the Participating Institutions, the National Aeronautics 
and Space Administration, the National Science Foundation, the U.S. Department of Energy, 
the Japanese Monbukagakusho, and the Max Planck Society. The SDSS Web site is http://www.sdss.org/. 
}
\vskip -0.4in
\phantom{x}

\end{document}